\documentstyle[12pt,graphicx]{article}
\begin{document}
\title{Vacuum Polarization for a Massless Scalar Field in the Global 
Monopole Spacetime at Finite Temperature}
\author{F. C. Carvalho\thanks{E-mail: fabiocc@ift.unesp.br}\\ and 
E. R. Bezerra de Mello \thanks{E-mail: emello@fisica.ufpb.br}\\
Departamento de F\'{\i}sica-CCEN\\
Universidade Federal da Para\'{\i}ba\\
58.059-970, J. Pessoa, PB\\
C. Postal 5.008\\
Brazil}
\maketitle
\begin{abstract}
In this paper we calculate the effects produced by the temperature in the 
renormalized vacuum expectation value of the square of the massless scalar 
field in the pointlike global monopole spacetime. In order to develop this 
calculation, we had to construct the Euclidean thermal Green function 
associated with this field in this background. We also calculate the 
high-temperature limit for the thermal average the zero-zero component of 
the energy-momentum tensor.
\end{abstract}

\newpage
\renewcommand{\thesection}{\arabic{section}.}
\section{Introduction}
$        $

It is well known that different types of topological defects may have  been 
created in the early Universe after the Planck time by the vacuum phase 
transition\cite{Kibble,Vilenkin}. These include domain walls, cosmic strings 
and monopoles. Among them cosmic string and monopole seem to be the best
candidates to be detected.

A global monopole is a heavy object formed in the phase transition of 
a system composed by a self-coupling scalar field triplet $\varphi^a$, whose 
original global $O(3)$ symmetry is spontaneously broken to U(1). The simplest 
model which gives rise to a global monopole, was presented by Barriola and 
Vilenkin \cite{Barriola}. The gravitational effects produced by this object 
may be approximated by a solid angle deficit in the (3+1)-dimensional 
spacetime whose line element is given by
\begin{equation}
\label{metric_mono}
ds^2=-\alpha^2dt^2+\frac{1}{\alpha^2}dr^2+r^2(\theta+\sin^2\theta 
d\varphi^2) \ .
\end{equation}
Here the parameter $\alpha^2=1-8\pi G\eta_0^2$ is smaller than unity and 
depends on the energy scale  $\eta_0$ where the global symmetry is 
spontaneously broken. The energy-momentum tensor of this monopole has a 
diagonal form and reads: $T^0_0=T^1_1=(\alpha^2-1)/r^2$ and $T^2_2=T^3_3=0$.

The non-trivial topology of this spacetime implies that the renormalized 
vacuum expectation value (VEV) of the energy-momentum tensor, 
$\langle T_{\mu\nu}(x)\rangle_{ren}$, associated with an arbitrary 
collection of conformal massless quantum fields should not 
vanish\cite{Hiscock}. The explicit calculations for the massless scalar
\cite{Mazzitelli} and fermionic\cite{Eugenio} fields have been already 
obtained.

In the framework of the quantum field theory at finite temperature, the 
fundamental quantity is the thermal Green function, $G_{\beta}(x,x')$. For the 
scalar field it should be periodic in the imaginary time with period $\beta$, 
which is proportional to the inverse of the temperature. Because we are 
interested to obtain the thermal Green function, it is convenient to work in 
the Euclidean analytic continuation of the Green function performing a Wick 
rotation. So, we shall work on the Euclidean version of the monopole metric,
\begin{equation}
\label{metric_Emono}
ds^2=d\tau^2+\frac{1}{\alpha^2}dr^2+r^2(\theta+\sin^2\theta d\varphi^2)\ ,
\end{equation}
where we have absorbed the $\alpha$-parameter redefining the Euclidean 
temporal coordinate.

The main objectives of this paper are to study the effects produced by the 
temperature on the renormalized VEV of the square of the massless scalar 
field operator and the respective energy-momentum tensor 
in the spacetime defined by (\ref{metric_Emono}). In order to do that, we 
calculate first the respective Euclidean thermal Green function in this 
manifold, which is the most relevant quantity for the obtaining of these 
expressions.

This paper is organized as follows. In section 2, we obtain the Euclidean 
thermal Green function $G_{\beta}(x,x')$ adopting the imaginary-time 
approach\cite{Dolan}, using the Schwinger-De Witt formalism for a massless 
scalar field. In section 3, we calculate the renormalized thermal 
average value $\langle\phi^2(x)\rangle_{\beta}$. Because this term cannot 
be written in a closed form, its dependence on the temperature is not 
evident. So, in order to obtain some quantitative information about its 
behavior, we have to proceed a numerical evaluation. In section 4, we present 
a formal expression for the thermal average of the zero-zero component of the 
energy-momentum tensor $\langle T_{00}(x)\rangle_{\beta}$. Again, 
because of its complicated dependence on the temperature, only numerical 
calculations enable us to present its appropriate behavior. In section 5, 
we present our conclusions and remarks about this paper. Finally, we 
left for the Appendix some details of our numerical calculations.

\section{The Euclidean Thermal Green Function}
$         $

The Euclidean Green function associated with a massless scalar field in 
the global monopole spacetime has been obtained by Mazzitelli and Lousto a few
years ago\cite{Mazzitelli}. There this Green function was obtained assuming 
a non-minimal coupling between the field and the geometry through the 
scalar curvature, $R=2(1-\alpha^2)/2r^2$, of this spacetime.

In this section we extend this result to obtain the Euclidean
Green function at finite 
temperature in the monopole spacetime defined by (\ref{metric_Emono}). The 
thermal Green function, $G_{\beta}(x,x')$, must obey the non-homogeneous 
Klein-Gordon differential equation, and be periodic in the "Euclidean" time 
$\tau$ with a period $\beta=1/\kappa_BT$, $\kappa_B$ being the Boltzmann 
constant and $T$ the absolute temperature. Considering the non-minimal 
coupling, the scalar Green function must obey the differential equation:
\begin{equation}
\label{KG_mono}
(\Box -\xi R)G_{\beta}(x,x')=-\delta^{(4)}(x,x') ,
\end{equation}
where $\Box$ denotes the covariant d'Alambertian operator in the metric 
defined by (\ref{metric_Emono}), $\xi$ is an arbitrary coupling constant and 
$\delta^{(4)}(x,x')$ is the bidensity Dirac distribution.

Because our spacetime is an ultrastatic one\footnote{An ultrastatic 
spacetime admits a globally defined coordinate system in which the components 
of the metric tensor are time independent and the conditions $g_{00}=1$ and  
$g_{0i}=0$ hold.}, the Euclidean thermal Green function can be obtained by the
Schwinger-De Witt formalism as follows:
\begin{equation}
\label{green_SDW}
G_{\beta}(x,x')=\int_0^{\infty}dsK_{\beta}(x,x';s)\ ,
\end{equation}
where the thermal heat kernel, $K_{\beta}(x,x';s)$, obeys the equation
\begin{equation}
\label{HK_mono}
\left(\frac{\partial}{\partial s}-\Box +\xi R\right)K_{\beta}(x,x',s)=0 
\,\,\,\, (s>0),
\end{equation}
is subject to the boudary condition
\begin{equation}
\label{BC}
\lim_{s\rightarrow0}K_{\beta}(x,x';s)=\delta^{(4)}(x,x') ,
\end{equation}
and periodic in the Euclidean time with period $\beta$.

Following the prescription given in the papers by Braden\cite{Braden} and 
Page\cite{Page}, the thermal heat kernel can be expressed in terms of the sum
\begin{equation}
\label{HK_sum}
K_{\beta}(x,x';s)=\sum_{n=-\infty}^{\infty}K_{\infty}(x,x'-n\lambda\beta;s)\ ,
\end{equation}
where $\lambda$ is the "Euclidean" time unit vector. The zero temperature heat 
kernel, $K_{\infty}(x,x';s)$, in this spacetime, can be factorized as 
\begin{equation}
\label{HK_T=0}
K_{\infty}(x,x';s)=K_{(1)}(\tau,\tau';s)K_{(3)}(\vec x,\vec x';s)\ ,
\end{equation}
where $K_{(1)}(\tau,\tau';s)$ and $K_{(3)}(\vec x,\vec x';s)$ obey,
respectively, the differential equations:
\begin{equation}
\label{HK_K1}
\left(\frac{\partial}{\partial s}-\frac{\partial^2}{\partial\tau^2}\right)
K_{(1)}(\tau,\tau';s)=0
\end{equation}
and
\begin{equation}
\label{HK_K3}
\left(\frac{\partial}{\partial s}-\nabla^2+\xi R\right)
K_{(3)}(\vec x,\vec x';s)=0\ .
\end{equation}
Here the covariant spatial Laplace operator has the form
\begin{eqnarray*}
\nabla^2=\frac{\alpha^2}{r^2}\partial_r(r^2\partial_r)-\frac{\vec L^2}{r^2}\ ,
\end{eqnarray*}
$\vec L^2$ is the square of the flat-space angular momentum operator.

Also, as was poited out in \cite{Camporesi}, the heat kernel can be 
expressed in terms of the eingenfunctions of the operators $-\Box+\xi R$ as 
\begin{equation}
\label{hk}
K(x,x';s)=\sum_{\sigma}\phi_{\sigma}(x)\phi_{\sigma}^{*}(x')e^{-s\sigma^2}\ ,
\end{equation}
$\sigma^2$ is the corresponding positively defined eigenvalues. Writing
\begin{equation}
\label{eq_eigen}
(-\Box+\xi R)\phi_{\sigma}(x)=\sigma^2\phi_{\sigma}(x)\ ,
\end{equation}
we get in our metrical spacetime the following results:
\begin{equation}
\label{eigenfunction}
\phi_{\sigma}(x)=\sqrt{\frac{\alpha p}{2\pi r}}Y_{l,m}(\theta,\varphi)
e^{-i\omega\tau}J_{\nu_{l}}(pr)
\end{equation}
and
\begin{equation}
\label{eingenvalue}
\sigma^2=\alpha^2p^2+\omega^2\ ,
\end{equation}
where $Y_{lm}(\theta,\phi)$ are the spherical harmonics, $J_{\nu_l}$ the 
Bessel functions and 
$\nu_{l}=\alpha^{-1}\sqrt{(l+1/2)^2+2(1-\alpha^2)(\xi-1/8)}$. 
So, according to (\ref{hk}) our zero temperature heat kernel is given by 
\begin{eqnarray}
\label{HK_infty}
K_{\infty}(x,x';s)&=&\int_{-\infty}^{\infty}\frac{d\omega}{2\pi}
\int_{0}^{\infty}dp\sum_{l,m}\phi_{\sigma}(x)\phi_{\sigma}(x')e^{-s\sigma^2}
\nonumber\\
&=&\frac{e^{-(\Delta\tau)^2/4s}}{2\sqrt{\pi s}}
\cdot\frac{e^{-\frac{r^2+r'^{2}}{4s\alpha^2}}}{8\pi s\alpha(rr')^{1/2}}
\sum_{l=0}^{\infty}(2l+1)I_{\nu_{l}}\left(\frac{rr'}{2s\alpha^2}\right)
P_l(\cos\gamma)\ ,
\nonumber\\
\end{eqnarray}
where we have used the addition theorem for the spherical harmonics, with
$\gamma$ satisfying the well known relation with the original angles 
$(\theta,\varphi)$ and $(\theta',\varphi')$. In (\ref{HK_infty}) 
$I_{\nu_l}$ is the modified Bessel function.

Comparing (\ref{HK_infty}) with (\ref{HK_T=0}) we can identify
\begin{equation}
\label{K1}
K_{(1)}(\tau,\tau';s)=\frac{e^{-(\tau-\tau')^2/4s}}{2\sqrt{\pi s}}
\end{equation}
and
\begin{equation}
\label{K3}
K_{(3)}(\vec x,\vec x';s)=\frac{e^{-\frac{r^2+r'^2}{4s\alpha^2}}}{8\pi s\alpha
(rr')^{1/2}}\sum_{l=0}^{\infty}(2l+1)I_{\nu_{l}}\left(\frac{rr'}{2s\alpha^2}
\right)P_l(\cos\gamma)\ .
\end{equation}

Now we want to call attention for the fact that for $\alpha=1$, $\nu_l$ become 
equal to $l+1/2$ and it is possible to get a closed expression for the sum 
in (\ref{K3})
\cite{Abramowitz}, getting the ordinary heat kernel function for the 
flat-space case
\begin{equation}
\label{K_infty_T=0}
K_{\infty}^{(\alpha=1)}(x,x';s)=\frac{1}{16\pi^2}\frac{e^{-(x-x')^2/4s}}
{s^2}\ .
\end{equation}

Now we are in position to calculate the heat kernel function at non-zero 
temperature in the global monopole spacetime. According to 
(\ref{HK_sum})-(\ref{HK_K1}) only $K_{(1)}(\tau,\tau';s)$ will be 
affected by temperature, so we get
\begin{equation}
\label{K1beta}
K_{(1)\beta}(\tau,\tau';s)=\frac{1}{2\sqrt{\pi s}}
\sum_{n=-\infty}^{\infty}e^{-(\tau-\tau'^2+n\beta)^2/4s}\ .
\end{equation}

Combining Eqs. (\ref{K1beta}) and (\ref{K3}), we have $K_{\beta}(x,x';s)$ 
and consequently our thermal Green function, $G_{\beta}(x,x')$, using 
(\ref{green_SDW}). Our result is:

\begin{equation}
\label{Gbeta}
G_{\beta}(x,x')=G_{\infty}(x,x')+\frac{1}{8\pi^2rr'}\sum_{n\neq0}
\sum_{l=0}^{\infty}(2l+1)Q_{\nu_l-1/2}(u_\beta)P_l(\cos\gamma)\ ,
\end{equation}
where 
\begin{equation}
\label{argument}
u_{\beta}=\frac{\alpha^2(\tau-\tau'+n\beta)^2+r^2+r'^2}{2rr'}\ ,
\end{equation}
and
\begin{equation}
\label{G_infty}
G_{\infty}(x,x')=\frac{1}{8\pi^2rr'}\sum_{l=0}^{\infty}(2l+1)
Q_{\nu_l-1/2}(u_{\infty})P_l(\cos\gamma)\ .
\end{equation}
(Unfortunately it is not possible to write this thermal Green function in 
terms of a single special function.)

Our temperature independent Green function $G_{\infty}(x,x')$ coincides, up 
to a redefinition of the "Euclidean" time $\tau$ by $\alpha\tau$, with the 
Green function for a massless scalar field at zero temperature given in 
Ref. \cite{Mazzitelli}.

For the case $\alpha=1$, Eq. (\ref{Gbeta}) provides
\begin{equation}
\label{G_alpha=1}
G_{\beta}^{(\alpha=1)}(x,x')=\frac{1}{4\pi\beta|\vec r-\vec r'|}
\frac{\sinh\left(2\pi/\beta|\vec r-\vec r|\right)}
{\cosh\left(\frac{2\pi}{\beta}|\vec r-\vec r'|\right)
-\cos\left(\frac{2\pi}{\beta}(\tau-\tau')\right)}\ .
\end{equation}
From the expression above it is possible to obtain its zero-temperature limit, 
taking $\beta\rightarrow\infty$. In this limit we get the ordinary Euclidean 
Green function given in the literature:
\begin{equation}
G(x,x')=\frac{1}{4\pi^2}\frac{1}{(x-x')^2}\ .
\end{equation}

\section{The Computation of $\langle\phi^2(x)\rangle_{\beta}$ at Nonzero 
Temperature}
$       $

The thermal average of $\langle\phi^2(x)\rangle_{\beta}$ can be obtained 
computing the coincidence limit of the Euclidean thermal Green function as shown below,
\begin{equation}
\label{phi2_def}
\langle\phi^2(x)\rangle_{\beta}=\lim_{x'\rightarrow x}G_{\beta}(x,x')
=\lim_{x'\rightarrow x}[G_{\infty}(x,x')+\overline{G}_{\beta}(x,x')],
\end{equation}
where we have separated the purely thermal part of (\ref{Gbeta}), defined as 
$\overline{G}_{\beta}(x,x')$.

The above procedure gives us a divergent result which comes exclusively 
from the zero-temperature contribuition of the thermal Green function, 
$G_{\infty}(x,x)$, which is proportional to the behavior of the Legendre 
function evaluated at the unity. So, in order to obtain a well
defined value for $\langle\phi^2(x)\rangle_{\beta}$, we extended to 
our thermal problem the renormalization procedure given in Ref. \cite{Wald}: 
we subtract from the complete thermal Green function the Hadamard function,
$G^{(1)}(x,x')$, which is given in terms of the square of the
biscalar geodesic interval, $\sigma(x,x')$. So, the renormalized
thermal average value for $\langle\phi^2(x)\rangle_{\beta}$ is given by 
\begin{equation}
\label{phi2_ren}
\langle\phi^2(x)\rangle_{\beta,Ren.}=\lim_{x'\rightarrow x}
[G_{\beta}(x,x')-G^{(1)}(x,x')]\ ,
\end{equation}
which is explicitly equal to
\begin{equation}
\label{phi2_ren2}
\langle\phi^2(x)\rangle_{\beta,Ren.}=\langle\phi^2(x)\rangle_{\infty,Ren.}
+\frac{1}{4\pi^2r^2}\sum_{n=1}^{\infty}\sum_{l=0}^{\infty}(2l+1)
Q_{\nu_l-1/2}\left(1+\alpha^2\frac{n^2\beta^2}{2r^2}\right).
\end{equation}

In their paper, Mazzitelli and Lousto\cite{Mazzitelli} have computed only the
zero temperature contribution, $\langle\phi^2(x)\rangle_{\infty,Ren.}$, up to
first order in the parameter $\eta^2=1-\alpha^2$, considered smaller than 
unity. So, we will not repeat their calculation. In this paper we compute the 
thermal contribution for the square of the scalar field operator up to
the first order in the parameter $\eta^2$ like in\cite{Mazzitelli}: 
\begin{equation}
\label{Phi-2}
\langle\overline{\phi}^2(x)\rangle_\beta=\frac{1}{4\pi^2r^2}
\sum_{n=1}^{\infty}\sum_{l=0}^{\infty}(2l+1)
Q_{\nu_l-1/2}\left( 1+\alpha^2\frac{n^2\beta^2}{2r^2}\right)\ .
\end{equation}
In order to do that let us take the integral 
representation for the Legendre function\cite{Gradstheyn}
\begin{equation}
\label{Qrepresent_int}
Q_{\nu-1/2}(\cosh\rho)=\frac{1}{\sqrt{2}}\int_{\rho}^{\infty}dt
\frac{e^{-\nu t}}{\sqrt{\cosh t-\cosh\rho}}\ .
\end{equation}
Substituting this expression in (\ref{Phi-2}), it is possible to proceed 
the summation on the angular quantum number, by using the expansion 
for $\nu_l$,
\begin{eqnarray*}
\nu_l\approx\left(l+\frac{1}{2}\right)\left(1+\frac{\eta}{2}\right)
+\frac{2\xi-1/4}{2l+1}\eta+\cdots
\end{eqnarray*}
After some other intermediate steps we obtain
\footnote{In the Appendix A we present in more details the steps adopted by us 
in this calculation.}
\begin{equation}
\label{phi2_barra}
\langle\overline{\phi}^2(x)\rangle_{\beta}=\frac{1}{12\beta^2}(1+\eta^2)
-\eta^2\xi\frac{\sqrt{2}}{8\pi^2r^2}S_1+\eta^2\frac{1}{32\pi^2r^2}S_2,
\end{equation}
where
\begin{equation}
\label{S1}
S_1=\sum_{n=1}^{\infty}\int_{\rho_n}^{\infty}dt\frac{1}{\sqrt{\cosh t
-\cosh\rho_n}}\frac{t}{\sinh(t/2)}
\end{equation}
and
\begin{equation}
\label{S2}
S_2=\sum_{n=1}^{\infty}\int_{\rho_n}^{\infty}dt\frac{1}{\sqrt{\cosh t
-\cosh\rho_n}}\frac{t}{\sinh^3(t/2)}
\end{equation}
with the notation $\rho_n=\mbox{arccosh}(1+n^2\beta^2/2r^2)$.

Unfortunately it was not possible to obtain an explicit result for both 
integrals above in terms of elementary or special functions. The best that we 
could do was to proceed a numerical evaluation of both integrals for specific 
values of the ratio $\zeta:=\beta/r$ and different values of $n$. In our 
numerical analyses the results were developed in the high temperature, or 
large distance limit, $\zeta\ll1$. They are shown in Figs.(1a) and 
(1b) for $S_1(\zeta)$ and  $S_2(\zeta)$ respectively. From the graphs displayed
in Fig. (2), which exhibit the logarithmic behavior for the previous 
figures, it is possible to infer the following dependences for 
$S_1(\zeta)$ and $S_2(\zeta)$ with $\zeta$: 
\begin{equation}
\label{S1_number}
S_1(\zeta)=\frac{c_1}{\zeta^{q_1}}
\end{equation}
and
\begin{equation}
\label{S2_number}
S_2(\zeta)=\frac{c_2}{\zeta^{q_2}}\ ,
\end{equation}
where $c_1=14.26$ with $q_1=0.92\pm0.07$ and  $c_2=17.82$ with 
$q_2=2.02\pm0.02$. From these results we can see that the most relevant 
contribution to $\langle\overline{\phi}^2\rangle_{\beta}$ is given by the term
independent on parameter $\xi$. It is, approximately, proportional to 
$1/\beta^2$ and consequently independent of the distance from the point to the
global monopole.

\section{The Thermal Average of $\langle T_{00}(x)\rangle_{\beta}$}
$        $

The energy-momentum tensor, $T_{\mu\nu}(x)$, is a bilinear function of the 
fields, so we can evaluate its vacuum expectation value, 
$\langle T_{\mu\nu}(x)\rangle_{\beta}$, by the standard method using the 
Green's function\cite{Birrel&Davis}. The thermal vacuum average, 
consequently, can also be obtained using the thermal Green function.

In Ref. \cite{Mazzitelli} the general structure of the
renormalized VEV of the energy-momentum tensor associated with a massless 
scalar field in the global monopole spacetime at zero temperature is presented.
(There, it was obtained considering the VEV 
of square of the field operator on the base of dimensional arguments,
symmetries and trace anomaly.) 
The objective of this section is to calculate the nonzero temperature 
correction to this vacuum polarization effect. So we concentrate 
on the thermal correction. The purely thermal average of 
energy-momentum tensor is given by:
\newpage
\begin{eqnarray}
\label{E-M}
\langle\overline{T}_{\mu\nu}(x)\rangle_{\beta}&=&\lim_{x'\rightarrow x}
\left[(1-2\xi)\nabla_{\mu}\nabla_{\nu'}\overline{G}_{\beta}(x,x')
-2\xi\nabla_{\rho}\nabla_{\nu}\overline{G}_{\beta}(x,x')\right.
\nonumber\\
&&\left.+\left(2\xi-\frac{1}{2}\right)g_{\mu\nu}(x)g^{\rho\sigma'}(x,x')
\nabla_{\rho}\nabla_{\sigma'}\overline{G}_{\beta}(x,x')\right.
\nonumber\\
&&\left.
-\xi G_{\mu\nu}(x)\overline{G}_{\beta}(x,x')
-2\xi^2g_{\mu\nu}R(x)\overline{G}_{\beta}(x,x')\right]\ ,
\nonumber\\
\end{eqnarray}
where $G_{\mu\nu}$ and $R$ are the Einstein tensor and the scalar 
curvature, respectively.

As we have already mentioned, we shall consider, for the sake of simplicity 
only, the zero-zero component of (\ref{E-M}); so, taking $\mu=\nu=0$ 
and using the fact that 
$\partial_{t'}\overline{G}(x,x')=-\partial_{t}\overline{G}(x,x')$, we obtain
\begin{eqnarray}
\langle\overline{T}_{00}(x)\rangle_{\beta}&=&\lim_{x'\rightarrow x}\left[
-\partial_{\tau}^2\overline{G}_{\beta}+\left(2\xi-\frac{1}{2}\right)
g_{\mu\nu}(x)g^{\rho\sigma'}(x,x')\nabla_{\rho}\nabla_{\sigma'}
\overline{G}_{\beta}(x,x')\right.
\nonumber\\
&&\left.+\frac{1}{2}\xi(1-4\xi)R(x)
\overline{G}_{\beta}(x,x')\right]\ .
\end{eqnarray}

On the other hand we know from (\ref{Gbeta}) that 
\begin{equation}
\label{Gthermal}
\overline{G}_{\beta}(x,x')=\frac{1}{8\pi^2rr'}\sum_{n\neq 0}^{\infty}
\sum_{l=0}^{\infty}(2l+1)Q_{\nu_l-1/2}\left(\frac{\alpha^2(\Delta\tau^2-
n\beta)^2+r^2+r^{'2}}{2rr'}\right)P_{l}(\cos\gamma).
\end{equation}
So, after some intermediate steps, we arrive at
\begin{eqnarray}
\label{stress0}
\langle\overline{T}_{00}(x)\rangle_{\beta}&=&\frac{\alpha^2}{4\pi^2 r^4}
\sum_{n=1}^{\infty}\sum_{l=0}^{\infty}(2l+1)[\Xi_n(\alpha,\zeta)
Q_{\nu_l-1/2}^{(2)}(z_n)
\nonumber\\
&&+\Phi_n(\alpha,\xi,\zeta)Q_{\nu_l-1/2}^{(1)}(z_n)
+\Psi_n(\alpha,\xi)Q_{\nu_l-1/2}(z_n)]\ ,
\end{eqnarray}
where
\begin{equation}
\label{Xi}
\Xi_n(\alpha,\zeta)=\frac{1}{(\alpha^2n^2\zeta^2+4)}[2\zeta(n^2\zeta^2-4)
-\frac{1}{2}(n^2\zeta^2+4)]\ ,
\end{equation}
\begin{eqnarray}
\label{Phi}
\Phi_n(\alpha,\xi,\zeta)&=&\frac{1}{n\zeta\alpha^3
\sqrt{(n^2\zeta^2+4)}}\left\{2\xi\left[(3n^2\zeta^2-2)
\alpha^2-2\right]\right. \nonumber\\
&&-\left.\frac{1}{2}\left[(3n^2\zeta^2-2)\alpha^2
-2\right]\right\}
\end{eqnarray}
and
\begin{equation}
\label{Psi}
\Psi_n(\alpha,\xi)=\left(2\xi-\frac{1}{2}\right)+\frac{l(l+1)}{\alpha^2}
+\frac{1}{2}\xi(1-4\xi)\frac{\alpha^2-1}{\alpha^2}.
\end{equation}
Here $z_n=1+\frac{n^2\alpha^2}{2}\zeta^2$ and $\zeta=\beta/r$. 
In Eq. (\ref{stress0}) $Q_{\nu}^{(m)}(z)$ are the associated Legendre 
functions given by
\begin{eqnarray*}
Q_{\nu}^{(m)}=(z^2-1)^{m/2}\frac{d^m}{dz^m}Q_{\nu}(z)
\end{eqnarray*}
for integer positive $m$.

The equations (\ref{stress0})-(\ref{Psi}) give a formal expression for the 
purely thermal correction to vacuum expectation value of 
$\langle T_{00}(x)\rangle_{\beta}$. However from these expressions it is not 
possible to obtain a concrete conclusion about its behavior with temperature. 
So, in order to provide this quantitative information, let us
expand $\langle\overline{T}_{00}\rangle_{\beta}$ in powers of
the parameter $\eta^2$ up to the first order. However, the dependence 
of $\langle\overline{T}_{00}(x)\rangle_{\beta}$ on $\eta^2=1-\alpha^2$ 
appears in two different ways: (i) in the coefficients (\ref{Xi})-(\ref{Psi}) 
and (ii) in the argument of the Legendre functions. 
After long calculations\footnote{In the Appendix B we present these 
calculations in more details.} which also include the summation in the angular 
quantum number $l$, we obtain an expression containing a large number of 
simple algebraic functions and integrals similar to the previous ones 
given in (\ref{S1}) and (\ref{S2}). Our final result is:
\begin{eqnarray}
\label{stress1}
\langle\overline{T}_{00}(x)\rangle_{\beta}&=&-\frac{\pi^2}{30\beta^4}
+\left(\frac{1}{24}-\xi\frac{1}{6}\right)\frac{1}{r^2\beta^2}+\left[\pi^2
\left(-\frac{11}{180}+\xi\frac{1}{9}\right)\frac{1}{\beta^4}\
+\right  .
\nonumber\\
&&\left.\Omega(\xi)\frac{1}{r^2\beta^2} +F(\xi,\zeta)\right]\eta^2\ ,
\end{eqnarray}
where
\begin{equation}
\Omega(\xi)=\frac{1}{24}(1-27\xi-\xi^2)\ .
\end{equation}
The function $F(\xi,\zeta)$ is expressed in terms of integrals as we show below
\begin{eqnarray}
F(\xi,\zeta)&=&\frac{1}{4\sqrt{2}\pi^2r^4}\left\{\left[
\frac{1}{4}f_{0,3}(\zeta)+\xi f_{0,1}(\zeta)
-\xi f_{2,3}(\zeta)-\xi^2 f_{2,1}(\zeta)\right]\right.
\nonumber\\
&&\left.-\left[\frac{1}{2}g_3(\zeta)
+\xi g_1(\zeta)\right]+\left(2\xi-\frac{1}{2}\right)
\left[\frac{1}{4}h_{4,1,5}(\zeta)
-\xi\frac{1}{2}h_{1,0,3}(\zeta)\right]\right\},
\nonumber\\
\end{eqnarray}
where
\begin{equation}
\label{ff}
f_{i,j}(\zeta)=\sum_{n=1}^{\infty}(3n^2\zeta^2-i^2)\left[
\frac{d}{dy}\int_{\mbox{arccosh}(y)}^{\infty}\frac{udu}{\sqrt{\cosh u-y}}
\frac{1}{\sinh^j(u/2)}\right]_{y=1+n^2\zeta^2/2},
\end{equation}
\begin{equation}
\label{gg}
g_{j}(\zeta)=\sum_{n=1}^{\infty}\overline{g}_n\,\overline{\lambda}_n^2\left[
\frac{d^2}{dy^2}\int_{\mbox{arccosh}(\large{y})}^{\infty}\frac{udu}
{\sqrt{\cosh u-y}}\frac{1}{\sinh^j(u/2)}\right]_{y=1+n^2\zeta^2/2},
\end{equation}
and
\begin{equation}
\label{hh}
h_{l,p,q}(\zeta)=\sum_{n=1}^{\infty}
\left[\int_{\mbox{arccosh}(y)}^{\infty}\frac{udu}{\sqrt{\cosh u-y}}
\frac{l^2+pe^{-u}}{\sinh^q(u/2)}\right]_{y=1+n^2\zeta^2/2},
\end{equation}
with $i=0,2$; $j=1,3$;  $l=1,2$; $p=0,1$ and $q=3,5$. We call attention for 
the fact that $\overline{g}_n=2\xi(n^2\zeta^2-4)-\frac{1}{2}
(n^2\zeta^2+4)$ and $\overline{\lambda}_n^2=n^2\zeta^2/2$.

Unfortunately, we could not express the integrals above in terms of elementary
functions. Once more, in order to provide a quantitative information about the
behavior of (\ref{stress1}), we have to proceed a numerical analysis of the 
integrals as a function of $\zeta=\beta/r$. However, the numerical 
calculations become more complicated here because of the derivatives which 
come from the development of $Q_{\nu}^{(1)}$ and $Q_{\nu}^{(2)}$. So making
an appropriate transformation of variable, $u:=\mbox{arccosh}\left(\frac
{y-1}q+1\right)$, the new integrals in the variable $q$ present fixed limits. 
This procedure allows us to take their derivative in a simple way. In the 
high temperature regime, $\zeta\ll1$, we have obtained the following 
numerical results:
\begin{eqnarray}
\label{stress2}
\langle\overline{T}_{00}(x)\rangle_{\beta}&=&-\frac{\pi^2}{30\beta^4}
+\left(\frac{1}{24}-\frac{\xi}6\right)\frac{1}{r^2\beta^2}+\left[
\Omega_1(\xi)\frac{1}{\beta^4}+\Omega_2(\xi)\frac{1}
{r^2\beta^2}\right.
\nonumber\\
&&\left.+F(\xi,\zeta)\right]\eta^2 \ ,
\end{eqnarray}
where
\begin{equation}
\Omega_1(\xi)=-\frac{\pi}{180}(11+52\xi) \ ,
\end{equation}
\begin{equation}
\Omega_2(\xi)=-\frac{1}{768}(64\xi^2+1725\xi-832)
\end{equation}
and
\begin{eqnarray}
\label{Ftilde}
F(\xi, \zeta)&=&\frac{1}{4\sqrt{2}\pi^2r^4}\left\{\left[
\frac{1}{4}\tilde{f}_{0,3}(\zeta)+\xi\tilde{f}_{0,1}(\zeta)
-\xi\tilde{f}_{2,3}(\zeta)-\xi^2\tilde{f}_{2,1}(\zeta)\right]\right.
\nonumber\\
&&\left.-\left[\frac{1}{2}\tilde{g}_3(\zeta)
+\xi\tilde{g}_1(\zeta)\right]+\left(2\xi-\frac{1}{2}\right)
\left[\frac{1}{4}\tilde{h}_{2,1,5}(\zeta)-
\frac{\xi}{2}\tilde{h}_{1,0,3}(\zeta)
\right]\right\}.
\nonumber\\
\end{eqnarray}
Here the behaviors of all the above expressions with $\zeta=\beta/r$ 
are given below \footnote{We exhibit only the contributions which depend on
$\zeta$ with power higher than or the same order as $O(1/ \zeta^2)$.}: \\
$(i)$ For the functions $\tilde{g}_j(\zeta)$ we have:
\begin{equation}
\tilde{g}_1=-\xi\left[\frac{132.71}{\zeta^{a_1}}+\frac{10.19}{\zeta^{b_1}}
+\frac{8.11}{\zeta^{c_1}}\right]+\frac{28.20}{\zeta^{a_2}}
+\frac{2.26}{\zeta^{b_2}}+\frac{8.11}{\zeta^{c_2}}
\end{equation}
and
\begin{equation}
\tilde{g}_3=-\xi\frac{129.84}{\zeta^{a_3}}+\frac{10.19}
{\zeta^{b_3}}\ ,
\end{equation}
with $a_1=2.26\pm0.04$, $b_1=2.03\pm0.03$, $c_1=3.00\pm0.00$ and
$a_2=2.22\pm0.02$, $b_2=1.96\pm0.23$, $c_2=2.99\pm0.01$. Also 
$a_3=4.01\pm0.01$ and $b_3=4.0\pm0.00$.\\
$(ii)$ For the functions $\tilde{f}_{ij}$ we have:
\begin{equation}
\tilde{f}_{03}=-\frac{55.21}{\zeta^{a_{4}}}
\end{equation}
with $a_{4}=1.96\pm0.01$,
\begin{equation}
\tilde{f}_{21}=-\frac{118.85}{\zeta^{a_{5}}}+\frac{21.27}{\zeta^{b_{5}}}
\end{equation}
with $a_{5}=2.27\pm0.07$ and  $b_{5}=3.04\pm0.03$,

\begin{equation}
\tilde{f}_{23}=-\frac{48.87}{\zeta^{a_{6}}}-
\frac{915.95}{\zeta^{b_{6}}} \ ,
\end{equation}
with $a_{6}=3.99\pm0.04$ and  $b_{6}=2.13\pm0.07$. \\
$(iii)$ Finally, for the functions $\tilde{h}_{l,p,q}(\zeta)$, we have:
\begin{equation}
\tilde{h}_{2,1,5}=\frac{77.63}{\zeta^{a_7}}
\end{equation}
and
\begin{equation}
\tilde{h}_{1,0,3}=\xi\frac{16.36}{\zeta^{b_7}}\ ,
\end{equation}
where $a_7=2.00\pm0.02$ and $b_7=3.97\pm0.03$.

From our numerical results, which were found for 
$\zeta=\beta/r$ belonging to the interval $[0.01, 0.1]$, we can 
conclude that most relevant terms are of order $1/\zeta^4$. According to 
(\ref{Ftilde}) these terms represent contributions to Eq. (\ref{stress1}) 
which are proportional to $1/\beta^4$ and are of the same order of the 
leading term, obtained analytically and independent of the distance. So the 
thermal average $\langle T_{00}(x)\rangle_{\beta}$ can be written in the form 
below, 
\begin{equation}
\langle T_{00}(x)\rangle^T_R=\langle T_{00}(x)\rangle^0_R+T^4g(\gamma),
\end{equation}
where $g$ is a function of the dimensionless variable $\gamma=rT$, whose
expression has been found by us approximately up to first order in the 
parameter $\eta^2$.

\section{Concluding Remarks}
$       $

In the present paper we have obtained the Euclidean thermal Green function 
$G_{\beta}(x,x')$ associated with a massless scalar field in a 
global monopole spacetime. Our result was expressed 
as a sum of a zero-temperature Green function plus a purely thermal 
contribution. Using this Green function we were able to obtain the 
corrections due to temperature in  
$\langle \phi^2(x)\rangle_{\beta}$ and $\langle T_{00}(x)\rangle_{\beta}$. 
Because in both calculations, the thermal contributions could not be expressed 
in a simple way, but some of them in terms of not solvable integrals, their
dependence on temperature could not be evaluated explicitly. So, in order
to get some quantitative information, we had to develop
a numerical analysis. We decided to consider these analyses for a 
high-temperature regime. Our final results for the thermal vacuum average 
$\langle\phi^2(x)\rangle_\beta$ and $\langle T_{00}\rangle_\beta$, 
present the following behavior:
\begin{equation}
\langle\phi^2(x)\rangle^T_R=\langle\phi^2(x)\rangle^0_R+T^2f(\gamma)
\end{equation}
and
\begin{equation}
\langle T_{00}(x)\rangle^T_R=\langle T_{00}(x)\rangle^0_R+T^4g(\gamma).
\end{equation}

For both case the functions $f$ and $g$, which depend on the specific geometry,
were found by us up to the first order in the parameter $\eta^2=1-\alpha^2$, 
considered smaller than unity. In fact for a typical grand unified theory 
the parameter $\eta_0$ is of order $10^{16}GeV$, so 
$\eta^2=8\pi\eta_0^2\approx10^{-5}$.

A few years ago, Smith\cite{Smith} and Linet\cite{Linet} have obtained the 
Euclidean thermal Green function and also the thermal average 
$\langle\phi^2(x)\rangle_{\beta}$ and $\langle T_{00}(x)\rangle_{\beta}$, for 
a massless scalar field in a cosmic string spacetime. For this case, Linet 
could present his results in a closed form. These two beautiful papers give us 
the motivation to analyse similar phenomena in a global monopole spacetime.

We want to emphasize that in all the above numerical evaluations, we 
considered sufficiently large number of terms in the summation  
in order to obtain good approximate results. (Independent numerical 
simulations provided us the minimal values for $n$ in each series.)

The obtained results may have some applications in the early cosmology, where 
the temperature of the Universe was really high. We can see that the thermal 
contributions to $\langle\phi^2(x)\rangle_{\beta}$ and 
$\langle T_{00}(x)\rangle_{\beta}$ modify significantly the zero temperature 
quantitaties in that epoch. Because of this, they should be taken into 
account, e.g., in the theory of structure formation.

\newpage
{\bf{Acknowledgments}}
\\       \\
We would like to thank Conselho Nacional de Desenvolvimento Cient\'\i fico e 
Tecnol\'ogico (CNpQ) and CAPES for partial financial support. We also thanks
N. R. Khusnutdinov and V. M. Mostepanenko for helpful assistance.

\renewcommand{\theequation}{\Alph{section}.\arabic{equation}}
\appendix
\section{High-temperature limit of the thermal average of 
$\langle\phi^2(x)\rangle_{\beta}$}
$        $

In this appendix we give a brief explanation of our procedure used to obtain 
the high-temperature limit of the thermal average of 
$\langle\phi^2(x)\rangle_{\beta}$. According to (\ref{phi2_ren2}), the purely 
thermal contribution to this quantity is:
\begin{equation}
\label{phi2_app}
\langle\overline{\phi}^2(x)\rangle_{\beta}=\frac{1}{4\pi^2r^2}
\sum_{n=1}^{\infty}\sum_{l=0}^{\infty}(2l+1)
Q_{\nu_l-1/2}(z_n-\eta^2\lambda_n^2)\ ,
\end{equation}
with
\begin{equation}
z_n=1+\frac{n^2\zeta}{2}\hskip1cm and\hskip1cm\lambda_n^2=
\frac{n^2\zeta^2}{2}\ ,
\end{equation}
where $\zeta=\beta/r$.

Expanding first the argument of the Legendre function in 
(\ref{phi2_app}) up to the first order in $\eta^2$, we get:
\begin{eqnarray}
\label{phi2_app2}
\langle\overline{\phi}^2(x)\rangle_{\beta}=\frac{1}{4\pi^2r^2}
\sum_{n=1}^{\infty}\sum_{l=0}^{\infty}(2l+1)\left[Q_{\nu_l}(z_n)
-\eta^2\lambda^2_n\left.\frac{d}{dy}Q_{\nu_l-1/2}(y)\right|_{y=z_n}\right].
\end{eqnarray}

We can substitute $\nu_l=l+1/2$ into the second
term of (\ref{phi2_app2}). This allows us to make the 
summation in the angular quantum number $l$ \cite{Gradstheyn}, and 
consequently make the second summation in $n$, resulting 
\begin{equation}
\label{Phi3}
\langle\phi^2(x)\rangle_{\beta}=\frac{1}{12\beta^2}\eta^2+\frac{1}{4\pi^2r^2}
\sum_{n=1}^{\infty}\sum_{l=0}^{\infty}(2l+1)Q_{\nu_l-1/2}(z_n).
\end{equation}

Using the integral representation (\ref{Qrepresent_int}) for the Legendre 
function and the expansion
\begin{eqnarray*}
\nu_l\approx\left(l+\frac{1}{2}\right)\left(1+\frac\eta 2\right)
+\frac{2\xi-1/4}{2l+1}\eta^2+\cdots\ ,
\end{eqnarray*}
we can obtain a simpler expression for (\ref{Phi3}) by the summation in 
$l$ as shown below:
\begin{equation}
\sum_{l=0}^{\infty}(2l+1)e^{-\nu_lt}=q^{1/2}\frac{1+q}{(1-q)^2}\left\{1
+2\frac{\eta^2\ln q}{(1-q^2)}[\xi(1-q)^2+q]\right\},
\end{equation}
where $q=e^{-t}$. (This summation was also present in Ref.\cite{Mazzitelli}. 
There is, however a misprint, the factor $\eta^2$  appears 
dividing by $2$, for this reason we reproduce the correct result.  
We want to call attention that this misprint, has been 
corrected in further calculations and no 
mistake was found by us). Now going back to (\ref{phi2_app2}) we get the 
following expression
\begin{equation}
\langle\overline{\phi}^2(x)\rangle_{\beta}=\frac{1}{12\beta^2}(1+\eta^2)
-\eta^2\frac{\xi\sqrt{2}}{8\pi^2r^2}S_1(\zeta)
+\eta^2\frac{1}{32\pi^2r^2}S_2(\zeta),
\end{equation}
where $S_1(\zeta)$ and $S_2(\zeta)$ are integrals given by 
(\ref{S1}) and (\ref{S2}) respectively.

\section{The expansion of (\ref{stress0}) in powers of $\eta^2$}
$       $

The expansion of (\ref{stress0}) in powers of $\eta^2$ is very long. 
In order to make our explanation clear to the reader, let us sepatate this 
calculations into three parts:\\
$(i)$ The first one refers to the term in $Q_{\nu_l-1/2}^{(2)}$, 
\begin{equation}
I_1=\frac{1}{4\pi^2r^4}(1-\eta^2)\sum_{n=1}^{\infty}\sum_{l=0}^{\infty}(2l+1)
\Xi_n(1-\eta^2,\xi,\zeta)Q_{\nu_l-1/2}^{(2)}(z_n-\eta^2\lambda_n^2).
\end{equation}

Making the expansion in the argument of the Legendre function in powers of 
$\eta^2$, we get:
\begin{eqnarray}
I_1&=&\frac{1}{4\pi^2r^4}(1-\eta^2)\sum_{n=1}^{\infty}\sum_{l=0}^{\infty}(2l+1)
\frac{\overline{\Xi}(\xi,\zeta)}{n^2\zeta^2+4}\left\{Q_{\nu_l-1/2}^{(2)}(z_n)
\right.
\nonumber\\
&&-\left.\left[\left(\frac{4}{n^2\zeta^2+4}\right)Q_l^{(2)}(z_n)
+\lambda_n^2\left.\frac{d}{dy}Q_l^{(2)}(y)\right|_{y=z_n}\right]
\eta^2\right\}\ ,
\end{eqnarray}
where
\begin{equation}
\overline{\Xi}_n(x,\zeta)=2\xi(n^2\zeta^2-4)-\frac{1}{2}
(n^2\zeta^2+4)\ .
\end{equation}

Using the definition for $Q_l^{(2)}$ and the integral representation for
the Legendre function, we obtain, after the expansion in powers of 
$\eta^2$ and the summation on $l$, the following expression:
\begin{eqnarray}
I_1&=&-\pi^2\frac{1}{45}(1+4\xi)\frac{1}{\beta^4}
-\frac{1}{12}(1+4\xi)\frac{1}{r^2\beta^2}+\left[-\pi^2\left(\frac{1}{30}
+\xi\frac{4}{45}\right)\frac{1}{\beta^4}\right.
\nonumber\\
&&\left.-\frac{1}{12}(1-4\xi)\frac{1}{r^2\beta^2}\right]\eta^2
+\frac{1}{4\sqrt{2}\pi^2r^4}\eta^2\left[\frac{1}{2}g_3(\zeta)
-\xi g_1(\zeta)\right],
\end{eqnarray}
where the functions $g_1(\zeta)$ and $g_2(\zeta)(z_n)$ are 
given in (\ref{gg}).

The other two terms of (\ref{stress0}) are developed in similar way. After 
some intermediate steps we obtain the following results:\\
$(ii)$ For the term proportional to $Q_{\nu_l-1/2}^{(1)}$ we have found
\begin{eqnarray}
I_2&=&\xi\frac{2\pi^2}{45}\frac{1}{\beta^4}+\frac{1}{8}
(1-4\xi)\frac{1}{r^2\beta^2}+\left[\pi^2\frac{1}{180}
(-1+20\xi)\frac{1}{\beta^4}+\frac{1}{8}
(1-12\xi)\frac{1}{r^2\beta^2}\right]\eta^2
\nonumber\\
&&+\frac{1}{4\sqrt{2}\pi^2r^4}\eta^2
\left[\frac{1}{4}f_{0,3}(\zeta)+\xi f_{0,1}(\zeta)-
\xi f_{2,3}(\zeta)-\xi^2 f_{2,1}(\zeta)\right]\ ,
\end{eqnarray}
where the functions $f_{0,3}(\zeta)$, $f_{0,1}(\zeta)$, 
$f_{2,3}(\zeta)$ and $f_{1,3}(\zeta)$ are given in  Eq. (\ref{ff}).\\
$(iii)$ For the term proportional to $Q_{\nu_l-1/2}(z_n)$ we have: 
\begin{eqnarray}
I_3&=&\frac{\pi^2}{90}(-1+4\xi)\frac{1}{\beta^4}
+\frac{1}{24}\xi(1-4\xi)\xi\frac{1}{r^2\beta^2}\eta^2
+\frac{\pi^2}{45}(-1+\xi)\frac{1}{\beta^4}\eta^2
\nonumber\\
&&+\frac{1}{4\sqrt{2}\pi^2r^4}\eta^2
\left[\frac{1}{4}h_{4,1,5}(\zeta)-
\xi\frac{1}{2}h_{1,0,3}(\zeta)\right]\ ,
\end{eqnarray}
where the functions $h_{2,1,5}(\zeta)$ and $h_{1,0,3}(\zeta)$ 
are given in the  
Eq. (\ref{hh}).


\begin{figure}[t]
\begin{center}
\includegraphics[width=10cm]{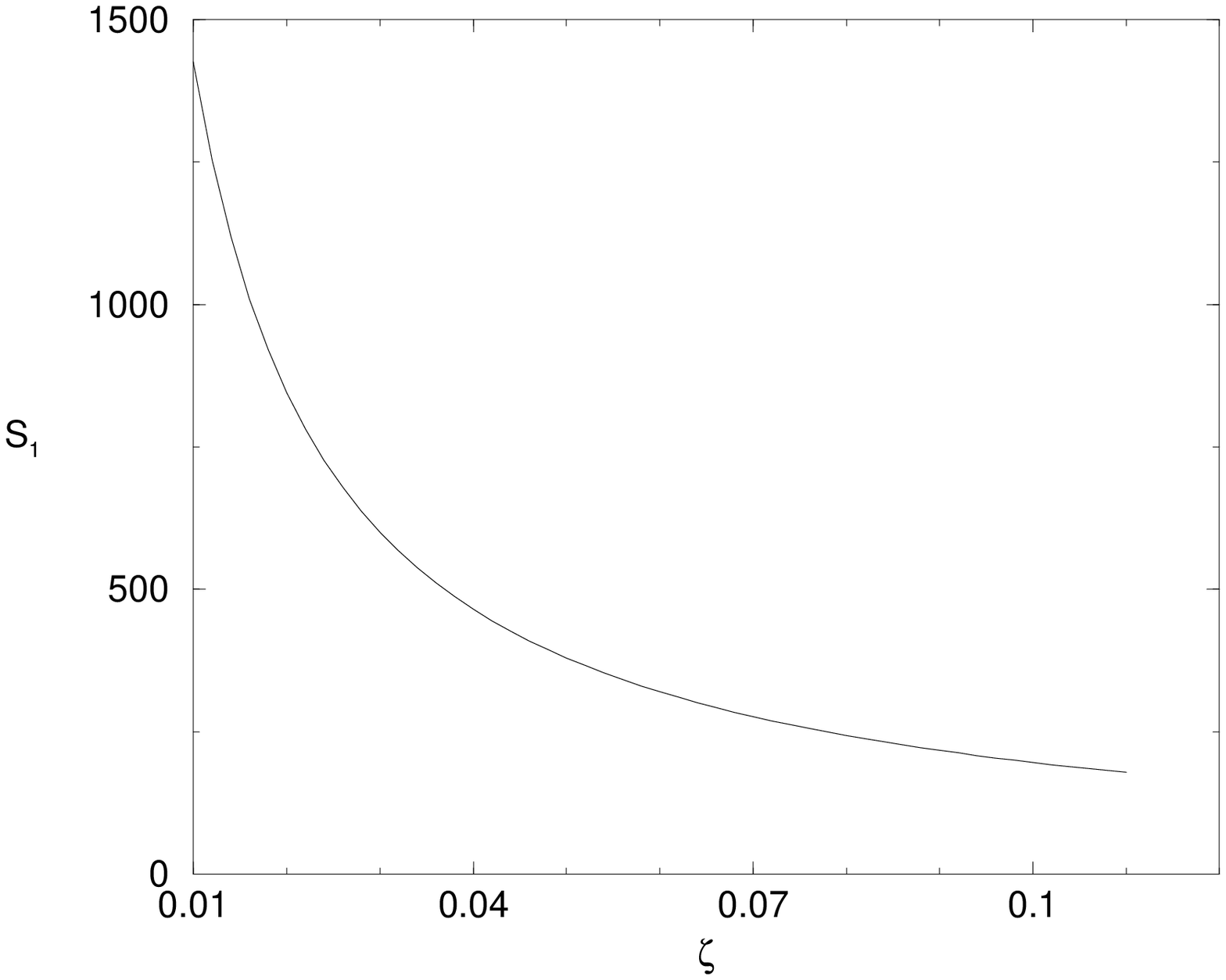}
%
\includegraphics[width=10cm]{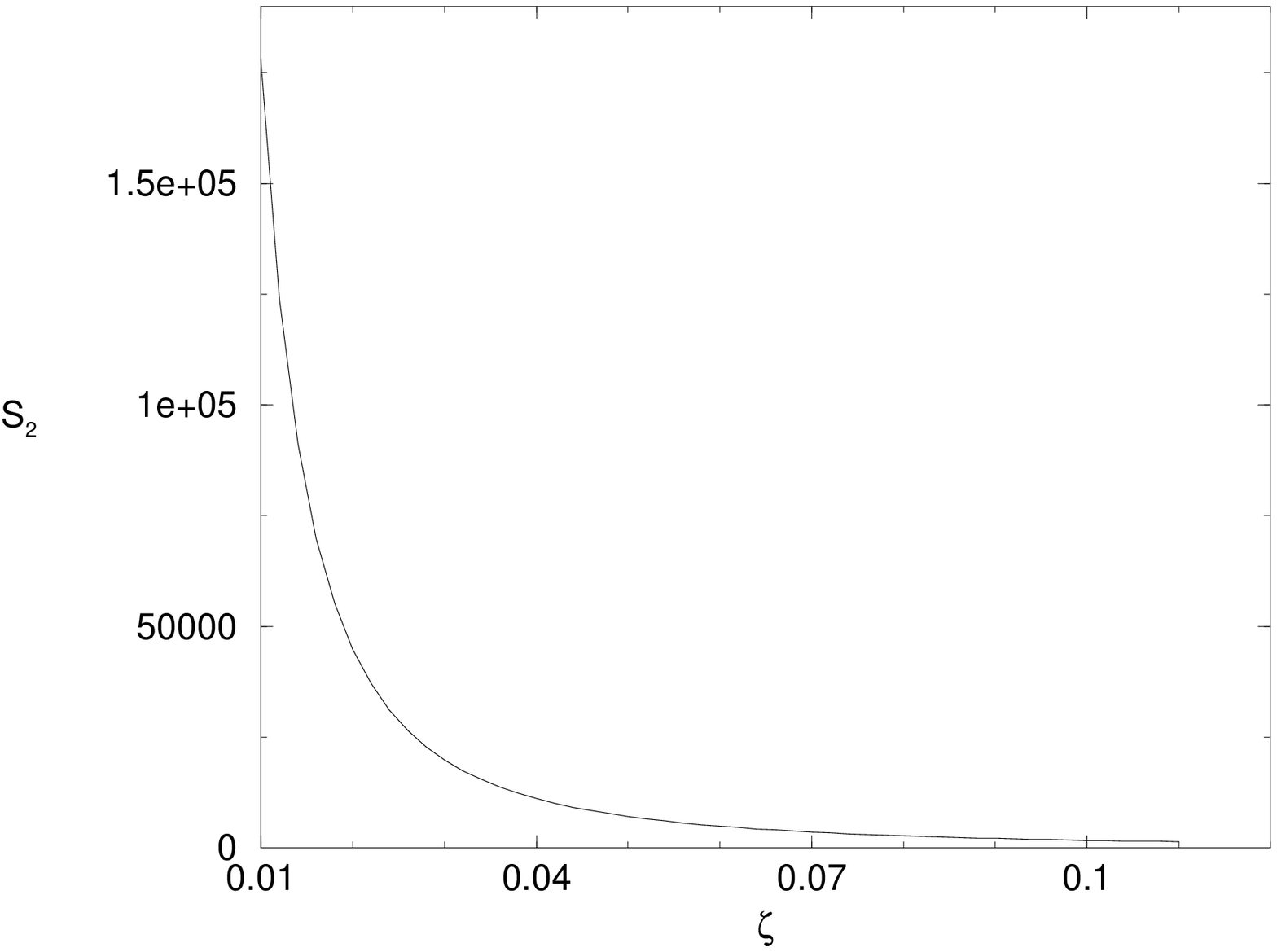}
\caption{These figures exhibit, respectively, the behavior of $S_1$ 
and $S_2$ with $\zeta$ in the region $[0.01, 0.1]$.}
\end{center}
\end{figure}

\begin{figure}[h]
\begin{center}
\includegraphics[width=10cm]{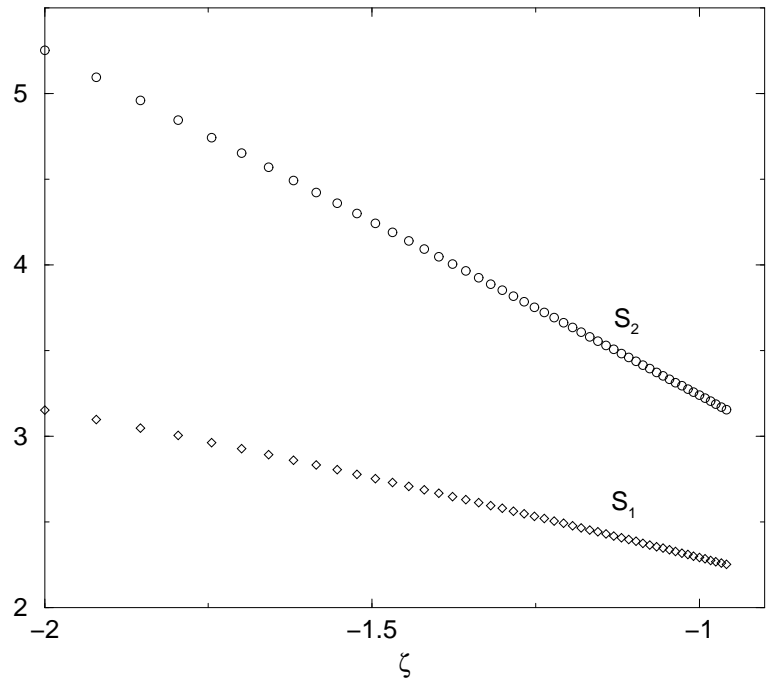}
\caption{This figure exibit the logaritmic behavior for both functions,
$S_1$ and $S_2$ with $\zeta$, in the region $[0.01,0.1]$. From it, it is
possible to estimate the leading dependence for both quantities with $\zeta$.}
\end{center}
\end{figure}

\end{document}